\documentclass[conference]{IEEEtran}
\IEEEoverridecommandlockouts
\usepackage{cite}
\usepackage{amsmath,amssymb,amsfonts}
\usepackage{algorithmic}
\usepackage{graphicx}
\usepackage{textcomp}
\usepackage{xcolor}
\usepackage{subfig}
\usepackage{hyperref}
\hypersetup{ 
    colorlinks=true,
    linkcolor=black,
    citecolor=black,      
    urlcolor=blue,
}

\def\BibTeX{{\rm B\kern-.05em{\sc i\kern-.025em b}\kern-.08em
    T\kern-.1667em\lower.7ex\hbox{E}\kern-.125emX}}
\begin{document}

\title{Automatic Recognition of Learning Resource Category in a Digital Library
}

\makeatletter
\newcommand{\linebreakand}{%
\end{@IEEEauthorhalign}
\hfill\mbox{}\par
\mbox{}\hfill\begin{@IEEEauthorhalign}
}
\makeatother

\author{\IEEEauthorblockN{Soumya Banerjee\IEEEauthorrefmark{1}, Debarshi Kumar Sanyal\IEEEauthorrefmark{2}, Samiran Chattopadhyay\IEEEauthorrefmark{3}, \\Plaban Kumar Bhowmick\IEEEauthorrefmark{4}, Partha Pratim Das\IEEEauthorrefmark{5}} \IEEEauthorblockA{\IEEEauthorrefmark{1}\IEEEauthorrefmark{4}\IEEEauthorrefmark{5}IIT Kharagpur, Kharagpur-721302, India, \IEEEauthorrefmark{2}Indian Association for the Cultivation of Science, Kolkata-700032, India, \\\IEEEauthorrefmark{1}\IEEEauthorrefmark{3}Jadavpur University, Kolkata-700106, India 
\\Email: \IEEEauthorrefmark{1}soumyaBanerjee@outlook.in,  \IEEEauthorrefmark{2}debarshisanyal@gmail.com, \IEEEauthorrefmark{3}samirancju@gmail.com,\\
\IEEEauthorrefmark{4}plaban@cet.iitkgp.ac.in,
\IEEEauthorrefmark{5}ppd@cse.iitkgp.ac.in}}

\maketitle

\begin{abstract}
	Digital libraries generally need to process a large volume of diverse document types. The collection and tagging of metadata is a long, error-prone, manpower-consuming task.  We are attempting to build an automatic metadata extractor for digital libraries. In this work, we present the Heterogeneous Learning Resources (HLR) dataset for document image classification. The individual learning resource is first decomposed into its constituent document images (sheets) which are then passed through an OCR tool to obtain the textual representation. The document image and its textual content are classified with state-of-the-art classifiers. Finally, the labels of the constituent document images are used to predict the label of the overall document.  
\end{abstract}


\section{Introduction}

A large digital library generally contains resources of different types. For example, the National Digital Library of India (NDLI) curates heterogeneous educational resources including scientific articles, books, paintings, etc. 
A library may receive curated metadata of resources directly or simply receive the resources from which it has to separately extract the metadata. 
In the latter case, knowing the type of the document is necessary because metadata extraction mechanisms (manual or automated) generally vary with resource types. 
Thus, it is worthwhile to explore methods of automatic classification of document types so that the correct metadata extraction process can be identified early. 

There is considerable literature on automatic classification of documents, using either textual information in the documents, or layout-specific information, or a combination of both. While most of the research in layout-based classification over the last three decades focused on ingenious hand-crafted features and rule-based or shallow machine learning algorithms \cite{chen2007survey}, the seminal publication by \cite{kang2014convolutional} in 2014 has sparked interest in the application of deep learning architectures to classify document images (see, e.g., \cite{harley2015evaluation}). 
However, recent approaches in existing literature mostly operate on the \textit{Tobacco} 
dataset which  has 3482 images from 10 classes or the larger more popular dataset \textit{RVL-CDIP} 
dataset containing 400K document images from 16 classes. The limitations of these approaches are that both datasets deal with single-page English textual documents. For metadata extraction in the context of a digital library, we need to process textual as well as non-textual multilingual multi-page document images. Moreover, the existing datasets do not adequately represent the content types in a typical educational digital library.
\begin{figure}
	\centering
	\includegraphics[width=0.90\linewidth]{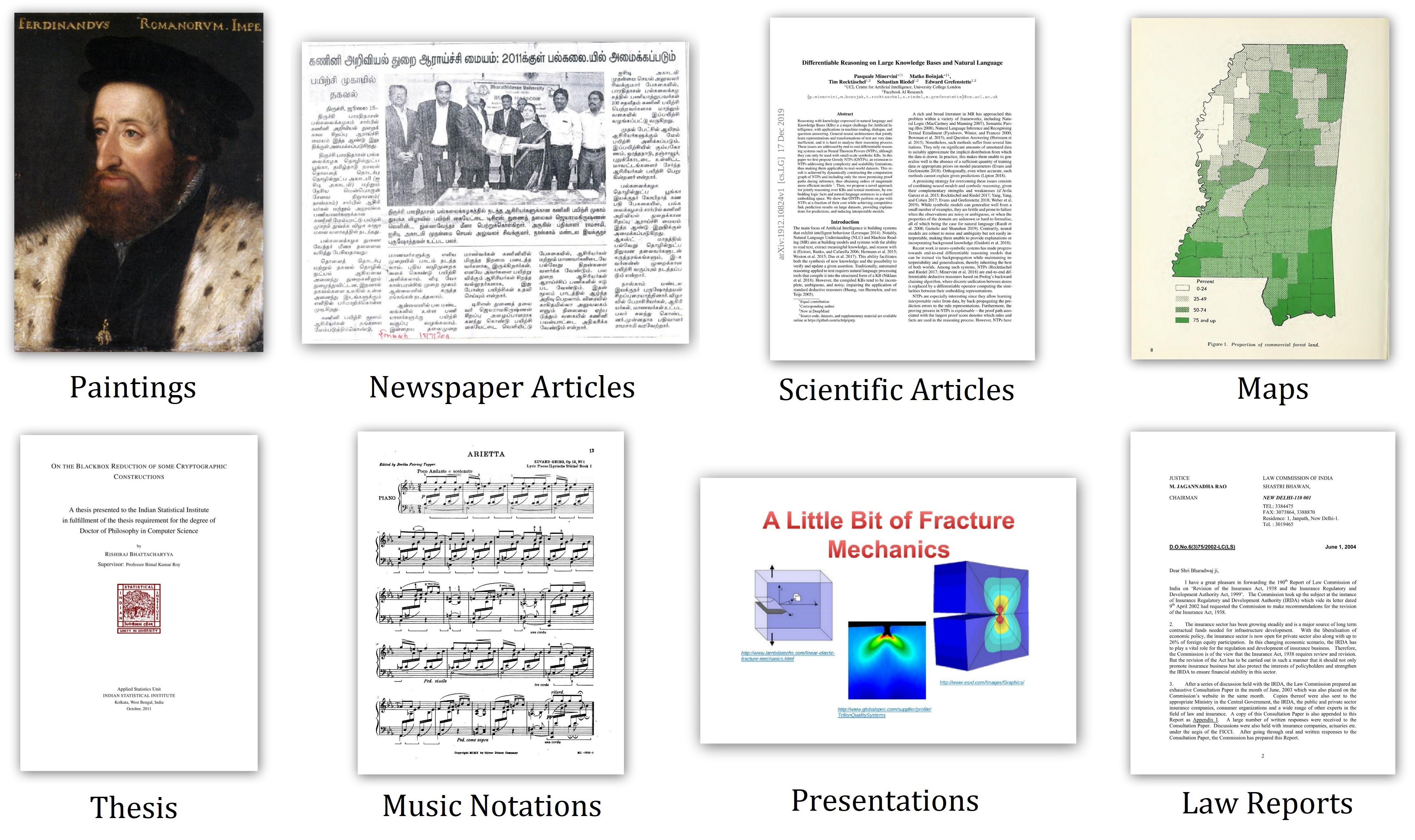}
	\caption{HLR Dataset Sample} \label{fig1}
\end{figure}
In this paper, we introduce the Heterogeneous Learning Resources (HLR) dataset to address the highlighted problem. We present benchmarks and demonstrate how existing techniques can be extended to address the said limitations.

\section{The HLR Dataset}
The proposed HLR dataset contains $3167$ images from $11$ classes, namely: catalog,  handwritten,  law reports,  maps,  music notations,  newspaper articles,  paintings,  presentation,  question paper,  scientific articles, and thesis.  
The data has been collected from NDLI and Europiana.
For document types with multiple pages, each page has been considered as independent samples. The dataset is split into an approximately $15:4:10$ training: validation: testing split. Additionally, a small set of multi-page documents are also included in the dataset for testing the said problem.  Figure \ref{fig1} shows a sample of the HLR dataset. The dataset and codes are available at  \url{https://github.com/soumyaxyz/DocumentClassify}.


\section{The Classifier Architecture}

We utilize a transfer learning-based training regime to train a classifier to identify document classes. This pre-trained classifier is employed to classify the multi-page documents. 
Architecturally, the deep learning model has two branches, the image subnetwork, and the textual subnetwork. The HLR dataset only contains images, with Tesseract-OCR, 
corresponding textual representations are generated. It is worth noting for a significant fraction of images, this textual representation is nothing but an empty string. Thus the textual subnetwork is strictly speaking an auxiliary network that assists the image subnetwork when possible.      
The image subnetwork the VGG16 architecture 
that is pre-trained on imagenet data as a feature extractor, the textual branch generates the GLOVE embeddings 
for the corresponding textual representation.  The embeddings are passed through a bi-LSTM with self-attention. Both are eventually concatenated and the final $11$ dimensional softmax is trained with an adam optimizer against a categorical cross-entropy loss function. 

The classifier, trained on the principal HLR dataset, achieves an impressive  $94.15$\% accuracy. 
Figure \ref{fig4} shows the confusion matrix for the classification task.
We also carried out a brief ablation study, where we investigated the subnetworks separately. The text-only branch failed to learn anything, this is quite expected as the textual data is very sparse. The image-only branch performed almost as well as the full model and achieved a $92.1$\% accuracy. But it performed worse in distinguishing between highly textual classes like Catalog, Law reports, and Scientific articles. 

\begin{figure}[!hbt]
	\centering
	\includegraphics[width=0.8\linewidth] {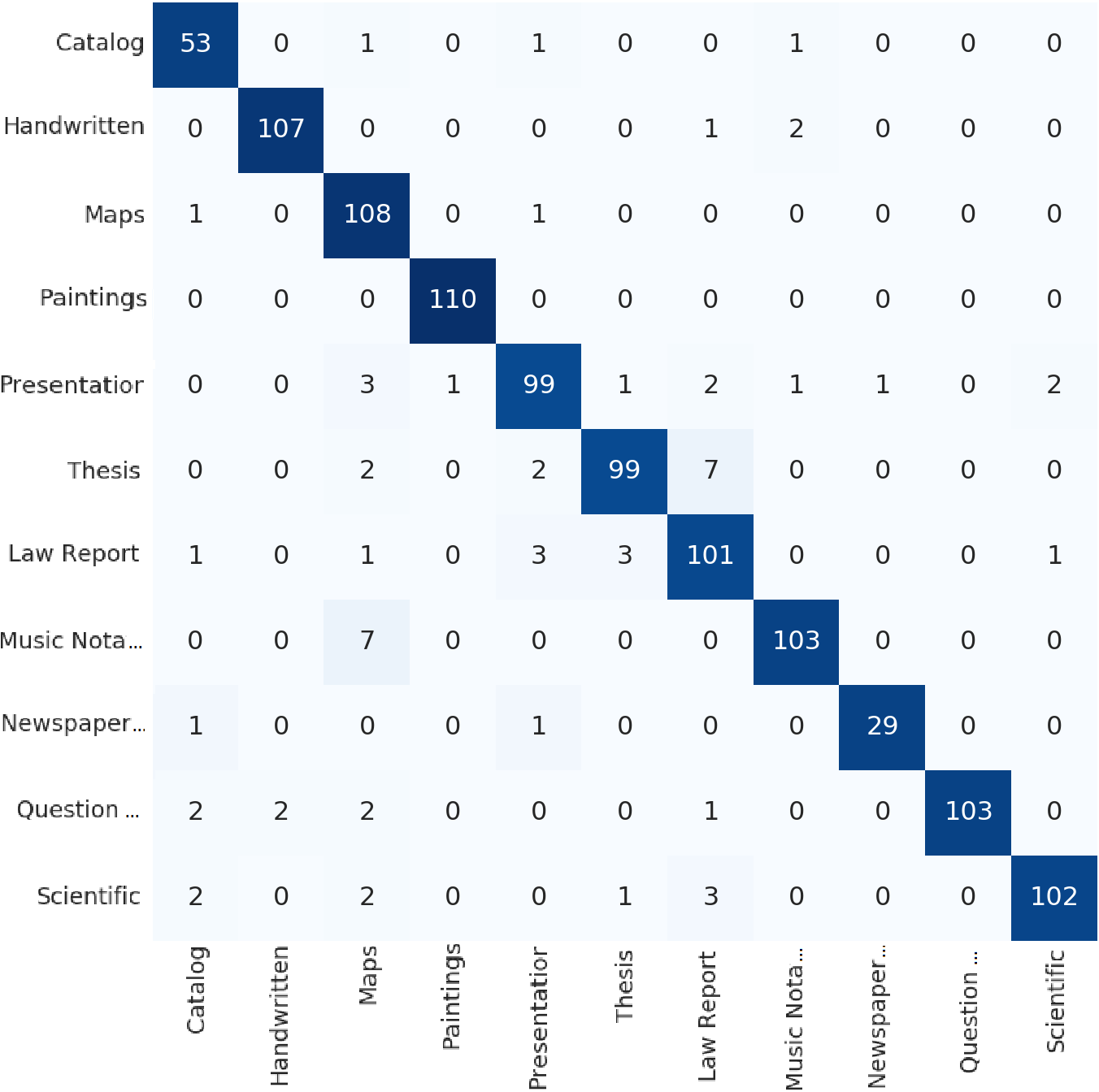}
	\caption{Confusion matrix for single page classification tasks.} \label{fig3}
\end{figure}

\vspace{-.2cm}

\section{Classification of Multi-Page Documents}
The HLR dataset is primarily a collection of single-page document images. But most of the classes (i.e., except handwritten, paintings, newspaper articles) are derived from multi-page documents which often contain images from multiple classes. 
It is easy for a human to identify the primary class for such documents. However, programmatically it is a nontrivial task. The HLR dataset also contains a small collection of multi-page documents comprising of $1483$ document images across $20$ documents. 

The trained classifier model is utilized to generate the labels for each page of these documents and the overall document class is predicted through a majority vote. This approach yields a respectable $80$\% accuracy. However, applying a bit of meta-knowledge about the dataset can significantly improve the performance. The \textit{map} documents have many tables and catalogs along with the titular maps, however, documents from the class \textit{catalog} do not contain any maps. Thus, this knowledge can be incorporated by checking the second label if the prediction is \textit{catalog} with low confidence.  If the second label is \textit{maps} with a comparable confidence, it should be reclassified as \textit{maps}. Similarly, \textit{thesis} and \textit{scientific articles} have very little distinction apart from the title page, thus \textit{scientific articles}  with the title page classified as \textit{thesis} should be reclassified as \textit{thesis}. Applying these corrections improves the accuracy to $95$\%. Figure \ref{fig4} shows the confusion matrices for the multi-page documents classifications, before and after the correction is applied.

\captionsetup[subfigure]{labelformat=empty}
\begin{figure}[!ht]
	\centering
	\subfloat[][\textit{Predicted}]{\includegraphics[width=.22\textwidth]{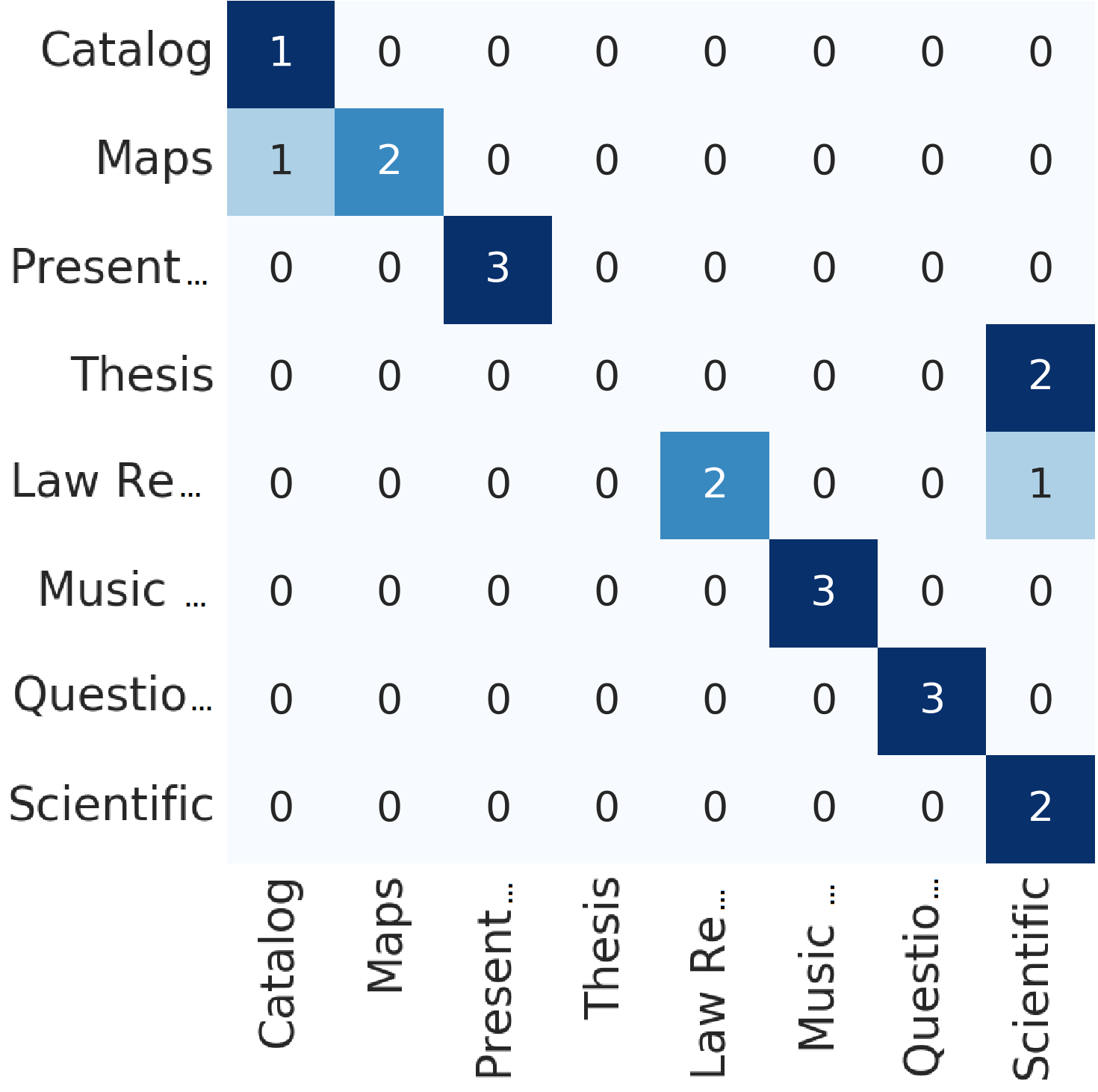}}
	\subfloat[][\textit{Corrected}]{\includegraphics[width=.22\textwidth]{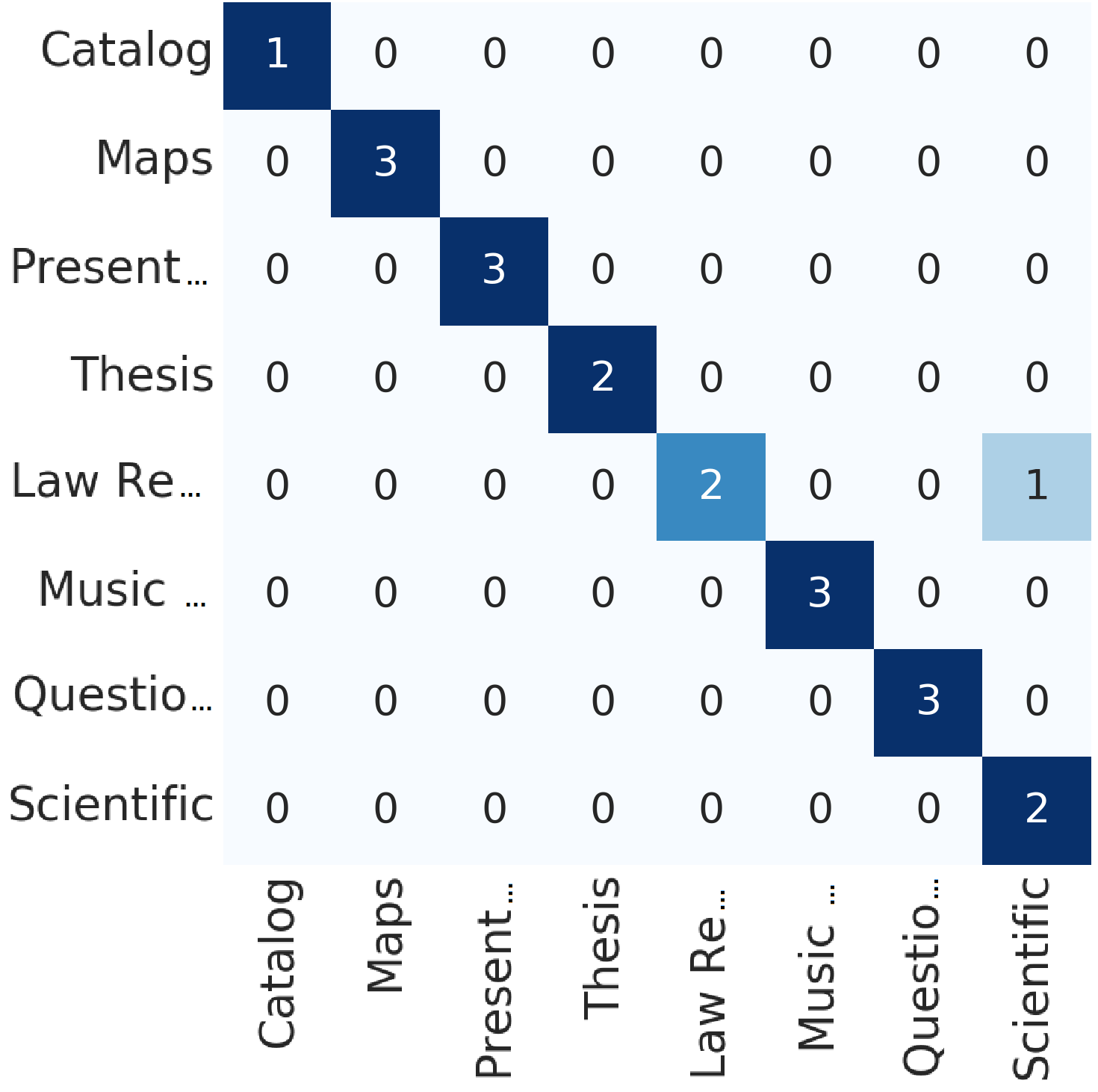}}
	\caption{Confusion matrices for multi-page classification tasks.}
	\label{fig4}
\end{figure}

\section{Conclusion}
We have presented a novel heterogeneous multi-lingual dataset for document image classification. We presented a deep-learning architecture for classifying heterogeneous document images. We also presented a system for on multi-page document classification. 
In the future, we will explore if the results generalize to larger multi-page datasets.

\section*{Acknowledgment}
This work is supported by the \textit{National Digital Library of India Project} sponsored by the Ministry of Education, Government of India at IIT Kharagpur.

\bibliographystyle{IEEEtran}
\bibliography{ref}

\end{document}